\documentclass[fleqn,10pt]{SelfArx}
\usepackage{lipsum}

\definecolor{color1}{RGB}{0,0,90} % Color of the article title and sections
\definecolor{color2}{RGB}{0,20,20} % Color of the boxes behind the abstract an

\usepackage{hyperref} % Required for hyperlinks
\hypersetup{hidelinks,colorlinks,breaklinks=true,urlcolor=color2,citecolor=color1,linkcolor=color1,bookmarksopen=false,pdftitle={Title},pdfauthor={Author}}

 \renewcommand{\vec}[1]{\mbox{\boldmath $#1$}}

 \def\aa{A\&A}
 \def\apj{ApJ}
 
 \def\an{Astron. Nachr.}

 \def\mnras{MNRAS}
 \def\nat{Nature}

\JournalInfo{Astronomy Letters, 2016, Vol. 42, No. 5} % Journal information
\Archive{}
\PaperTitle{
Rotational shear near the solar surface as a probe for subphotospheric magnetic fields
}
\Authors{L.~L.~Kitchatinov\textsuperscript{1,2}*}
\affiliation{\textsuperscript{1}\textit{Institute for Solar-Terrestrial Physics, Lermontov Str. 126A, Irkutsk, 664033, Russia}}
\affiliation{\textsuperscript{2}\textit{
Pulkovo Astronomical Observatory, Pulkovskoe Sh. 65, St. Petersburg, 196140, Russia
}}
\affiliation{*\textbf{E-mail}: kit@iszf.irk.ru}

\newcommand{\keywordname}{Keywords}

\Abstract{
Helioseismology revealed an increase in the rotation rate with depth just beneath the solar surface. The relative magnitude of the radial shear is almost constant with latitude. This rotational state can be interpreted as a consequence of two conditions characteristic of the near-surface convection: the smallness of convective turnover time in comparison with the rotation period and absence of a horizontal preferred direction of convection anisotropy. The latter condition is violated in the presence of a magnetic field. This raises the question of whether the subphotospheric fields can be probed with measurements of near-surface rotational shear. The shear is shown to be weakly sensitive to magnetic fields but can serve as a probe for sufficiently strong fields of the order of one kilogauss. It is suggested that the radial differential rotation in extended convective envelopes of red giants is of the same origin as the near-surface rotational shear of the Sun.
}

\begin{document}

\flushbottom % Makes all text pages the same height

\maketitle % Print the title and abstract box

%\tableofcontents % Print the contents section

\thispagestyle{empty} % Removes page numbering from the first page

%----------------------------------------------------------------------------------------
%	ARTICLE CONTENTS
%----------------------------------------------------------------------------------------
%%%%%%%%%%%%%%%%%%%%%%%%%%%%%%%%%%%%%%%%%%%%%%%%%%%%%%%%%%%%%%%%%%%%%%%%%%%%%
\section{Introduction} % The \section*{} command stops section numbering
%%%%%%%%%%%%%%%%%%%%%%%%%%%%%%%%%%%%%%%%%%%%%%%%%%%%%%%%%%%%%%%%%%%%%%%%%%%%%
Helioseismology revealed large inhomogeneity of solar rotation near both  boundaries of the convection zone. There is a shallow tachocline just beneath the base of the convection zone where the latitudinal differential rotation rapidly decreases with depth to converge to a uniform rotation of the radiation core (Antia et al. 1998; Charbonneau et al. 1999). Another relatively thin ($\sim 30$\,Mm) layer with rotation rate sharply increasing with depth lies right beneath the solar surface (Thompson et al. 1996). This paper concerns the surface shear layer.

It is remarkable that the relative rotational shear,\\ $\partial \log (\Omega)/\partial \ln (r) = -1$, in the surface layer is almost constant from the equator to 60$^\circ$ latitude (Ba\-re\-kat et al. 2014) though the angular velocity $\Omega$ and its gradient in (heliocentric) radius $r$ are latitude-dependent. More specifically, the relative shear averaged over 10\,Mm depth below the photosphere is uniform with latitude.

We shall see that the uniform surface shear can easily be explained in the framework of the differential rotation theory. It is a consequence of the two conditions characteristic of the near-surface region: 1) the characteristic time $\tau$ of turbulent convection is small compared to the rotation period, and 2) the only preferred direction of the turbulence anisotropy is the radial one.

The suggested explanation is as follows. As has been first shown by Lebedinskii (1941), anisotropic turbulence pro\-du\-c\-es the so-called \lq non-diffusive' fluxes of angular momentum, which are proportional to the angular velocity itself rather than to its gradient. This phenomenon is now called the $\Lambda$-effect (R\"udiger 1989). There are also diffusive fluxes of angular momentum, which are proportional to the gradient of the angular velocity, due to the turbulent viscosity. The standard boundary conditions, considered in the next Section, demand the total flux of angular momentum to be zero near the stellar surface. In other words, the viscosity and the $\Lambda$-effect balance each other at the surface. In the case of small turnover time $\tau \ll \Omega^{-1}$ and radial anisotropy of the turbulence, this balance demands the relative radial shear to be constant irrespective of how complex the dependencies on latitude in angular velocity and its gradient are.

The above explanation is however valid for weak magnetic fields only. A sufficiently strong field provides an additional anisotropy that involves a dependence on latitude in the relative rotational shear. This raises the question of whether helioseismological data similar to that obtained by Barekat et al. (2014) can provide information on subsurface solar magnetic fields? The question seems to be topical because the near-surface layer is considered as a possible site of the solar dynamo because of its large rotational shear (Brandenburg 2005; Pipin \& Kosovichev 2011). In view of the lack of measurements of internal solar magnetic fields, any means for the fields detection are significant.

In this paper, the $\Lambda$-effect is derived and the relative surface rotational shear is estimated with allowance for the magnetic field. We shall see that the shear is weakly sensitive to the magnetic field. The shear can, nevertheless, serve as a probe for a magnetic field whose toroidal component reaches the value of the order of one thousand Gauss.
%%%%%%%%%%%%%%%%%%%%%%%%%%%%%%%%%%%%%%%%%%%%%%%%%%%%%%%%%%%%%%%%%%%%%%%%%%%%%
\section{Continuity equation for the angular momentum and boundary conditions}
%%%%%%%%%%%%%%%%%%%%%%%%%%%%%%%%%%%%%%%%%%%%%%%%%%%%%%%%%%%%%%%%%%%%%%%%%%%%%
Rotational shear near the surface of the convective envelope of a star can be inferred from the boundary conditions which in turn follow from the continuity equation for angular momentum.

The angular momentum equation can be derived by averaging the equation for the fluid velocity $\vec v$. The velocity $\vec{v} = \vec{V} + \vec{u}$ of a fluid with convective turbulence includes the mean large-scale part $\vec{V}$ and the small-scale fluctuations $\vec{u}$. The averaging (over an ensemble of turbulent flows) separates the scales: $\langle\vec{v}\rangle = \vec{V}$, $\langle\vec{u}\rangle = 0$, where the angular brackets signify the averaging. Similarly, the magnetic field is a superposition of the mean large-scale field, $\vec B$, and fluctuating small-scale field, $\vec b$. In what follows, spherical coordinates ($r,\theta ,\phi$) with the axis of rotation as the polar axis are used and axial symmetry of the averaged stellar parameters is assumed. The continuity equation for the angular momentum,
\begin{eqnarray}
    \rho r^2\sin^2\theta \frac{\partial \Omega}{\partial t} =
    - \mathrm{div}{\big(} r\sin\theta( \rho\langle u_\phi\vec{u}\rangle -
    \langle b_\phi\vec{b}\rangle/4\pi &&
    \nonumber \\
    +\ \rho r\sin\theta\Omega\vec{V} - B_\phi\vec{B}/4\pi){\big)}\ ,&&
    \label{1}
\end{eqnarray}
results from the averaging of the azimuthal component of the motion equation (R\"udiger 1989). The vector under the divergence sign on the right-hand side of this equation is the angular momentum flux.

The standard boundary condition for Eq.\,(\ref{1}) is the continuity of radial flux of angular momentum on the surface of a star. The condition can be obtained by integrating Eq.\,(\ref{1}) over a small volume comprising a part of the surface. On taking into account that the mass flux $\rho V_r$ on the boundary is small (small mass loss for the stellar wind), that the large-scale magnetic field is continuous, and that convective turbulence is present inside the star only, we find the boundary condition
\begin{equation}
    T_{r\phi} = -\rho\langle u_ru_\phi\rangle + \langle b_r b_\phi\rangle/4\pi = 0 .
    \label{2}
\end{equation}
This condition for the off-diagonal component of the stress tensor, $T_{ij} = \langle -\rho u_iu_j + (b_ib_j - \delta_{ij}b^2/2)/4\pi\rangle$, means that the surface density of azimuthal forces is zero and the differential rotation is controlled by internal processes in the convection zone, not by an external impact. The condition (\ref{2}) is formulated for the surface but holds validity for depths that are small compared to the scale of variation of the stress tensor.

\begin{figure}\centering
\includegraphics[width=\linewidth]{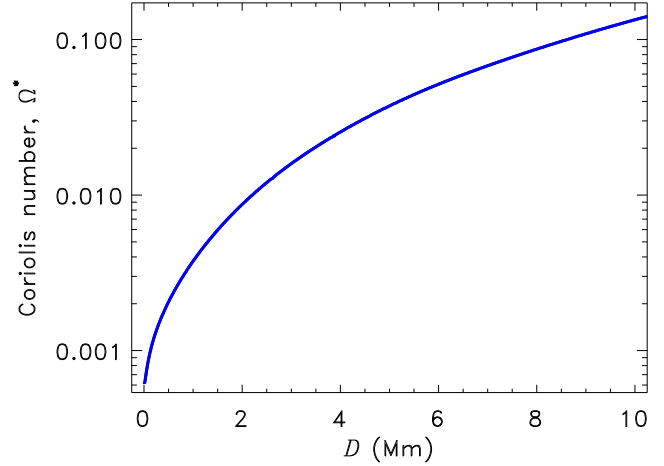}
\caption{Dependence of the Coriolis number (\ref{4}) on the
    depth $D$ beneath the solar surface.}
    \label{f1}
\end{figure}

The stress-tensor for rotating fluids with anisotropic turbulence includes two distinct components: non-diffusive str\-e\-s\-s\-es $T^\Lambda_{ij}$ and the contribution of turbulent viscosities $T^D_{ij}$,
\begin{equation}
    T_{ij} = T^\Lambda_{ij} + T^D_{ij},\ \ \ T^D_{ij} = \rho {\cal N}_{ijkl}\frac{\partial V_k}{\partial r_l} ,
    \label{3}
\end{equation}
where ${\cal N}_{ijkl}$ is the eddy viscosity tensor. Non-diffusive stress is the main source for the differential rotation (the $\Lambda$-effect), while the viscous stress $T^D_{ij}$ constrains the magnitude of rotation inhomogeneity by the action of turbulent viscosities (Kitchatinov 2005). Both the $\Lambda$-effect and the turbulent viscosities depend on rotation rate in a complicated way. The key parameter for the dependence is the Coriolis number
\begin{equation}
    \Omega^* = 2\tau\Omega
    \label{4}
\end{equation}
($\Omega^*$ measures the intensity of interaction between convection and rotation). The dependence of $\Omega^*$ on depth near the solar surface is shown in Fig.\,\ref{f1}. This dependence follows from the solar structure model (Stix 1989) and the value of $\Omega = 2.87\times 10^{-6}$s$^{-1}$ for the angular velocity. The Figure shows that $\Omega^*$ is small near the surface.

In the case of small $\Omega^*$ and without magnetic fields, relatively simple expressions hold for the stress tensor components present in Eq.\,(\ref{3}) (R\"udiger 1989),
\begin{eqnarray}
    T^\Lambda_{r\phi} &=& -S\rho\nu_{_\mathrm{T}}\Omega\sin\theta,
    \nonumber \\
    T^D_{r\phi} &=& \rho\nu_{_\mathrm{T}}r\sin\theta\frac{\partial\Omega}{\partial r} ,
    \label{5}
\end{eqnarray}
where $\nu_{_\mathrm{T}}$ is the turbulent viscosity and $S$ is the convection anisotropy parameter, which will be defined latter. Equation (\ref{5}) omits the contributions of the order of ${\Omega^*}^2$ the relative magnitude of which does not exceed 1\% (Fig.\,\ref{f1}). The boundary condition (\ref{2}) with account for Eq.\,(\ref{5}) leads to the constant relative radial shear,
\begin{equation}
    \frac{\partial \ln (\Omega )}{\partial \ln (r)} = S ,
    \label{6}
\end{equation}
near the solar surface (Kitchatinov 2013). This finding agrees with the helioseismological data of Barekat et al. (2014) and their value of the rotational shear restricts the anisotropy parameter $S$. However, Eq.\,(\ref{6}) is modified by a magnetic field.
%%%%%%%%%%%%%%%%%%%%%%%%%%%%%%%%%%%%%%%%%%%%%%%%%%%%%%%%%%%%%%%%%%%%%%%%%%%%%
\section{Lambda-effect in presence of a magnetic field}
%%%%%%%%%%%%%%%%%%%%%%%%%%%%%%%%%%%%%%%%%%%%%%%%%%%%%%%%%%%%%%%%%%%%%%%%%%%%%
\subsection{Quasi-linear approximation}
%%%%%%%%%%%%%%%%%%%%%%%%%%%%%%%%%%%%%%%%%%%%%%%%%%%%%%%%%%%%%%%%%%%%%%%%%%%%%
Turbulent stresses will be derived in the quasi-linear approximation also known as the first-order smoothing (Moffatt 1978) or the second-order correlation approximation (Krause \& R\"a\-dl\-er 1980). In this approximation, the terms that are nonlinear in turbulent fluctuations are neglected in the equations for the fluctuating velocity or magnetic field. The approximation allows the correlations of velocity or magnetic fields, similar to that present in (\ref{2}), to be derived as functions of rotation rate and/or mean magnetic field. The quasi-linear approximation is justified for cases when the Strouhal number or the  Reynolds number is small. Solar convective turbulence does not belong to either of these cases. Nevertheless, the presence of special cases for which the quasi-linear approximation is valid ensures this approximation from nonphysical results. Numerical experiments of K\"apyl\"a \& Brandenburg (2008) show the quasi-linear theory to keep the \lq order of magnitude validity' also beyond these special cases.

The quasi-linear technique for deriving the $\Lambda$-effect has been discussed in detail elsewhere (cf., e.g., Kitchatinov et al. 1994a) and is described here only briefly. It is convenient to apply the Fourier transform of fluctuating velocity,
\begin{equation}
    {\vec u}({\vec r},t) = \int \hat{\vec u}({\vec k},\omega )
    \mathrm{e}^{i({\vec r}\cdot{\vec k} - \omega t)}d{\vec k}d\omega ,
    \label{7}
\end{equation}
and the same for magnetic fluctuations $\vec b$. For the uniform angular velocity $\Omega$ and the mean field $\vec B$, this results in algebraic equations for the Fourier-amplitudes of the fluctuating fields, which lead to the following relations
\begin{eqnarray}
    \hat{u}_i(\vec{k},\omega ) &=&
    D_{ij}(\vec{k},\omega , \vec{\Omega},\vec{V_{_\mathrm{A}}})
    \hat{u}^0_j(\vec{k},\omega ),
    \nonumber \\
    \hat{\vec b}(\vec{k},\omega) &=&
    \frac{i(\vec{k}\cdot\vec{B})}{\eta k^2 - i\omega}\hat{\vec u}(\vec{k},\omega ) ,
    \label{8}
\end{eqnarray}
where $\vec{V_{_\mathrm{A}}} = \vec{B}/\sqrt{4\pi\rho}$ is the Alfven velocity, repetition of subscripts means summation, and $\hat{\vec u}^0$ corresponds to the so-called \lq original turbulence' which is not perturbed by rotation or magnetic field. The influence of rotation and magnetic field is involved via the tensor $D_{ij}$,
\begin{eqnarray}
    D_{ij}(\vec{k},\omega , \vec{\Omega},\vec{V_{_\mathrm{A}}}) =
    \frac{N\delta_{ij} + \hat{\Omega}\varepsilon_{ijl}k_l/k}{N^2 + \hat{\Omega}^2} ,\hspace{2.1 truecm} &&
    \nonumber \\
    N = 1 + \frac{(\vec{k}\cdot\vec{V_{_\mathrm{A}}})^2}
    {(\eta k^2 - i\omega)(\nu k^2 + i\omega)} ,\ \ \
    \hat{\Omega} = \frac{2(\vec{k}\cdot\vec{\Omega})}{k(\nu k^2 - i\omega)} ,&&
    \label{9}
\end{eqnarray}
where $\nu$ is the viscosity and $\eta$ is the magnetic diffusivity. The original turbulence is assumed to be given. The turbulent stresses $T^\Lambda_{ij}$ are derived from the given properties of the original turbulence by using equations (\ref{8}) and (\ref{9}) to account for the influence of rotation and magnetic field. The original turbulence is assumed to be anisotropic with the radial preferred direction signified by the unit vector $\hat{\vec r} = \vec{r}/r$. Properties of such a turbulence are defined by the spectral tensor
\begin{eqnarray}
    \hat{Q}_{ij}(\vec{k},\omega ) = \frac{E(k,\omega)}{16\pi k^4}
    \bigg[ (1 + S)( k^2\delta_{ij} - k_ik_j ) \hspace{1.2 truecm} &&
    \nonumber \\
    -\ S\left( (\hat{\vec r}\cdot\vec{k})^2\delta_{ij}
    + k^2\hat{r}_i\hat{r}_j
    - (\hat{\vec r}\cdot\vec{k})(\hat{r}_ik_j + \hat{r}_jk_i)\right)\bigg]&&
    \label{10}
\end{eqnarray}
of the velocity correlation
\begin{equation}
    \langle \hat{u}^0_i(\vec{k},\omega) \hat{u}^0_j(\vec{k}',\omega')\rangle =
    \hat{Q}_{ij}(\vec{k},\omega ) \delta(\vec{k} + \vec{k}')\delta(\omega + \omega') .
    \label{11}
\end{equation}
In this equation, $E(k,\omega)$ is the fluctuation spectrum,
\begin{equation}
    3\langle (u^0_r)^2\rangle = \int\limits_{0}^{\infty}\int\limits_{0}^{\infty}
    E(k,\omega)\ dk\,d\omega ,
    \label{12}
\end{equation}
and $S$ is the anisotropy parameter,
\begin{equation}
    S = \frac{\langle (u^0)^2\rangle}{\langle (u^0_r)^2\rangle} -3 ,
    \label{13}
\end{equation}
which has been used already in equations (\ref{5}) and (\ref{6}). The spectral tensor should be positive definite. This condition imposes restrictions on the anisotropy parameter:
\begin{equation}
    S \geq -1 .
    \label{14}
\end{equation}
According to (\ref{14}), the turbulent flow cannot consist of only radial motions (it would be $S = -2$ in this case). The mean energy density of horizontal motions cannot be smaller than that of radial motions.

The $\Lambda$-effect with allowance for the magnetic field can be derived from equations (\ref{8}) to (\ref{11}) by performing the inverse Fourier transform.
%%%%%%%%%%%%%%%%%%%%%%%%%%%%%%%%%%%%%%%%%%%%%%%%%%%%%%%%%%%%%%%%%%%%%%%%%%%%%%
\subsection{The $\Lambda$-effect}
\begin{figure}\centering
\includegraphics[width=7.4 truecm]{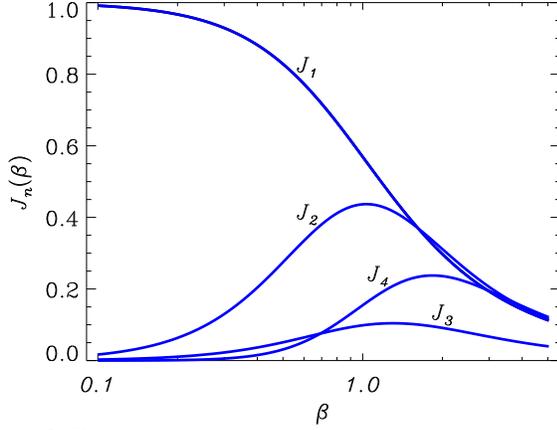}
\caption{Dependence of the $J_n$-coefficients of equation (\ref{15})
    on the relative strength $\beta$ (\ref{18}) of the large-scale magnetic field.}
    \label{f2}
\end{figure}

%%%%%%%%%%%%%%%%%%%%%%%%%%%%%%%%%%%%%%%%%%%%%%%%%%%%%%%%%%%%%%%%%%%%%%%%%%%%%%
The near-surface convection is weakly affected by rotation (Fig.\,\ref{f1}). It, therefore, suffices to derive linear terms in the angular velocity only in the $\Lambda$-effect. This yields
\begin{eqnarray}
    T^\Lambda_{ij} = S\rho\nu_{_\mathrm{T}} \bigg[ J_1
    \left( \hat{r}_i\varepsilon_{jmn} + \hat{r}_j\varepsilon_{imn}\right)
    \hat{r}_m\Omega_n
    \hspace{2.0 truecm} &&
    \nonumber \\
    -\ J_2(\vec{\Omega}\cdot\hat{\vec b})
    \left( \hat{r}_i\varepsilon_{jmn} + \hat{r}_j\varepsilon_{imn}\right)
    \hat{r}_m\hat{b}_n
    \hspace{2.0 truecm} &&
    \nonumber \\
    + J_3(\hat{\vec r}\cdot\hat{\vec b})
    \left( \Omega_i\varepsilon_{jmn} + \Omega_j\varepsilon_{imn}\right)
    \hat{r}_m\hat{b}_n
    \hspace{2.0 truecm} &&
    \nonumber \\
    + J_3 \left( \hat{b}_i\varepsilon_{jmn} + \hat{b}_j\varepsilon_{imn}\right)
    \left((\vec{\Omega}\cdot\hat{\vec r})\hat{r}_m\hat{b}_n
    +(\hat{\vec r}\cdot\hat{\vec b})\hat{r}_m\Omega_n\right)  &&
    \nonumber \\
    - J_4(\vec{\Omega}\cdot\hat{\vec b})(\hat{\vec r}\cdot\hat{\vec b})
    \left( \hat{b}_i\varepsilon_{jmn} + \hat{b}_j\varepsilon_{imn}\right)
    \hat{r}_m\hat{b}_n \bigg] ,
    \hspace{0.8 truecm} &&
    \label{15}
\end{eqnarray}
where $\hat{\vec b} = \vec{B}/B$ is the unit vector along the mean field $\vec B$ and the $J_n$-coefficients ($n=1,...,4$) depend on the field strength and on the fluid parameters. These dependencies are expressed in terms of integrals of the turbulence spectrum with complicated weight-functions. Further simplifications are required for quantitative estimations. Sufficient simplifications are provided by the mixing-length approximation (known as the $\tau$-approximation also). In this approximation, the spectrum function contains the only spatial scale $\ell = \tau u$ ($u = \langle (u^0)^2 \rangle^{1/2}$ is the RMS velocity):
\begin{equation}
    E(k,\omega ) = 2 u^2 \delta(k - \ell^{-1})\delta(\omega) .
    \label{16}
\end{equation}
Viscosity and magnetic diffusivity are replaced by their estimated effective values,
\begin{equation}
    \nu = \eta = \ell^2/\tau .
    \label{17}
\end{equation}
The $J_n$-coefficients of Eq.\,(\ref{15}) then depend only on the ratio
\begin{equation}
    \beta = B/B_\mathrm{eq},\ \ \ B_\mathrm{eq} = \sqrt{4\pi\rho}\ u ,
    \label{18}
\end{equation}
of the mean field $B$ to the so-called equipartition field $B_\mathrm{eq}$ (for $B = B_\mathrm{eq}$, the magnetic energy density equals the density of turbulent kinetic energy). For small $\beta$, only $J_1$ tends to a finite value different from zero: $J_1 \simeq 1 - 6\beta^2/7$, $J_2 \simeq 12\beta^2/7$, $J_3 \simeq 2\beta^2/7$ and $J_4 = O(\beta^4)$ (for $\beta \ll 1$). In the opposite case of a very strong field, all $J_n$-coefficients are inversely proportional to the field strength: $J_1 \simeq J_2 \simeq J_4 \simeq 45\pi/(256\beta)$ and $J_3 \simeq 15\pi/(256\beta)$ (for $\beta \gg 1$). The $J_n$-functions for intermediate $\beta$-values are shown in Fig.\,\ref{f2}.

%%%%%%%%%%%%%%%%%%%%%%%%%%%%%%%%%%%%%%%%%%%%%%%%%%%%%%%%%%%%%%%%%%%%%%%%%%%%%
\section{Near-surface rotational shear}
\begin{figure}\centering
\includegraphics[width=7.8 truecm]{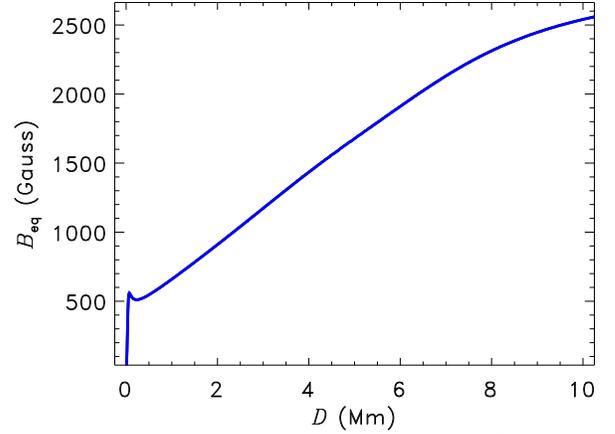}
\caption{Dependence of the equipartition field $B_\mathrm{eq}$ on depth $D$
    beneath the solar surface.}
    \label{f3}
\end{figure}

%%%%%%%%%%%%%%%%%%%%%%%%%%%%%%%%%%%%%%%%%%%%%%%%%%%%%%%%%%%%%%%%%%%%%%%%%%%%%
We turn now to the estimation of the near-surface radial inhomogeneity of rotation with allowance for the magnetic field. The field can change considerably the angular momentum fluxes only if it is not too small compared with $B_\mathrm{eq}$ ($B > 0.1 B_\mathrm{eq}$, Fig.\,\ref{f2}). Figure \ref{f3} shows the dependence of the equipartition field $B_\mathrm{eq}$ (\ref{18}) on depth beneath the solar surface for the same solar structure model as Fig.\,\ref{f1}. It follows from Fig.\,\ref{f3} that only the fields of at least several hundred Gauss can influence the near-surface rotational shear. The large-scale poloidal field is much weaker, $B^\mathrm{pol} \sim 1$\,Gs (Stenflo 1988; Obridko et al. 2006). The toroidal field of the solar $\alpha\Omega$-dynamo can be much stronger. Further estimations are, therefore, performed for toroidal field $\vec{B} = \vec{e}_\phi B$, that largely simplifies the estimations.

Equation (\ref{15}) leads to the following expression for the non-dissipative part $T^\Lambda_{r\phi}$ (\ref{3}) of the radial flux of angular momentum,
\begin{equation}
    T^\Lambda_{r\phi} = -S\rho\nu_{_\mathrm{T}}\Omega J_1(\beta)\sin\theta ,
    \label{19}
\end{equation}
which generalises the Eq.\,(\ref{5})$_1$ with allowance for the magnetic field.

\begin{figure}\centering
\includegraphics[width=7.8 truecm]{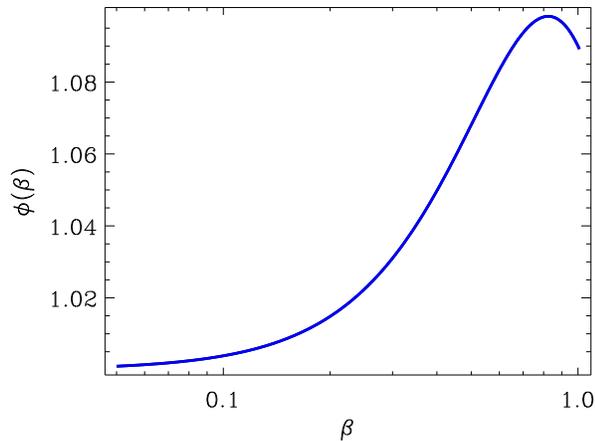}
\caption{The function $\phi(\beta)$ of the dependence of the rotational shear
    on the toroidal field strength near the surface of a stellar convection zone.}
    \label{f4}
\end{figure}

Estimating the rotation inhomogeneity requires a knowledge of the viscous stresses which balance the $\Lambda$-effect. The viscosity tensor of equation (\ref{3}), derived within the same approximations as the $\Lambda$-effect (\ref{15}), reads
\begin{eqnarray}
    {\cal N}_{ijkl} = \nu_{_\mathrm{T}}\big{[} \psi_1(\beta)(\delta_{ik}\delta_{jl} + \delta_{jk}\delta_{il})
    \hspace{2.7 truecm} &&
    \nonumber \\
    \ +\ \psi_2(\beta)(\delta_{il}\hat{b}_j\hat{b}_k
    + \delta_{jl}\hat{b}_i\hat{b}_k + \delta_{ik}\hat{b}_j\hat{b}_l
    + \delta_{jk}\hat{b}_i\hat{b}_l ) + ... \big{]}, &&
    \label{20}
\end{eqnarray}
where dots signify the terms irrelevant to this paper (the functions $\psi_1(\beta)$ and $\psi_2(\beta)$ are given in the Appendix of the paper of Kitchatinov et al. (1994b) and in its figure 4). The contribution of the eddy viscosity to the radial flux of angular momentum then reads
\begin{equation}
    T^D_{r\phi} =\rho\nu_{_\mathrm{T}} \left( \psi_1(\beta) + \psi_2(\beta)\right)
    r\sin\theta\frac{\partial\Omega}{\partial r} .
    \label{21}
\end{equation}
Substitution of the stresses (\ref{19}) and (\ref{21}) into the boundary condition (\ref{2}) finally gives
\begin{equation}
    \frac{\partial\ln(\Omega)}{\partial\ln(r)} = S\phi(\beta),
    \label{22}
\end{equation}
where $\phi(\beta) = J_1(\beta)/(\psi_1(\beta) + \psi_2(\beta))$ is the function of the normalised strength of the toroidal magnetic field (\ref{18}). This function is shown in Fig.\,\ref{f4}.

%%%%%%%%%%%%%%%%%%%%%%%%%%%%%%%%%%%%%%%%%%%%%%%%%%%%%%%%%%%%%%%%%%%%%%%%%%%%%
\section{Conclusion}
%%%%%%%%%%%%%%%%%%%%%%%%%%%%%%%%%%%%%%%%%%%%%%%%%%%%%%%%%%%%%%%%%%%%%%%%%%%%%
Without magnetic fields, the shear (\ref{22}) is constant with latitude. The shear value found by Barekat et al. (2014) is reproduced with the maximum possible radial anisotropy $S = -1$ (\ref{14}), i.e, with $\langle u_r^2\rangle = 2\langle u_\theta^2\rangle = 2\langle u_\phi^2\rangle$. Numerical experiments of K\"apyl\"a et al. (2011) show that the anisotropy of non-rotating convection is indeed close to this maximum value (see fig.\,5 of K\"apyl\"a et al.).

For $S = -1$, the normalized rotational shear equals the $\phi(\beta)$-function of Fig.\,\ref{f4}: $-\partial\ln(\Omega)/\partial\ln(r) = \phi(\beta)$. The shear increases with the field strength. The increase is caused by the stronger suppression of turbulent viscosity by the magnetic field compared with the $\Lambda$-effect. The magnetic field decreases the turbulence intensity. As a consequence, both the $\Lambda$-effect and the eddy viscosity balancing this effect are decreased. However, the viscosity is more strongly quenched that results in an increase in the rotational shear with the magnetic field (Fig.\,\ref{f4}). As the toroidal field depends on latitude, a similar dependence can emerge in the near-surface rotational shear. Figure\,2 of Barekat et al. (2014) does indeed show a small increase in the rotational shear around the latitude of 20$^\circ$ (see also table\,1 in their paper). However, the relatively small value of this increase does not allow its confident interpretation as a manifestation of a subphotospheric magnetic field. A comparison of Figures \ref{f3} and \ref{f4} shows that the surface rotational shear can serve as an indicator of sufficiently strong subphotospheric magnetic fields of the order of one thousand Gauss.

The theory of the solar surface shear layer can be relevant to the rotational states of red giants. Asteroseismology has revealed a large inhomogeneity of rotation in extended convective envelopes of these stars (Beck et al. 2012; Deheuvels et al. 2012). The relatively fast rotation of their deep interiors was interpreted as a consequence of angular momentum conservation in the course of contraction of the central region and expansion of the envelope of a star. The characteristic time of turbulent diffusion ($\sim100$~yers) in the convection zone of a solar-mass giant is however short compared with its evolutionary time. This circumstance together with long rotation periods suggests that the quasi-steady non-uniform rotation in extended convective envelopes of red giants can be of the same nature as the near-surface rotational shear of the Sun.
%%%%%%%%%%%%%%%%%%%%%%%%%%%%%%%%%%%%%%%%%%%%%%%%%%%%%%%%%%%%%%%%%%%%%%%%%%%%%
\phantomsection
\section*{Acknowledgments}
%\addcontentsline{toc}{section}{Acknowledgments} % Uncomment to add Acknowledgements to the table of contents
%%%%%%%%%%%%%%%%%%%%%%%%%%%%%%%%%%%%%%%%%%%%%%%%%%%%%%%%%%%%%%%%%%%%%%%%%%%%%
This work was supported by the Russian Foundation
for Basic Research (project 16--02--00090).
%%%%%%%%%%%%%%%%%%%%%%%%%%%%%%%%%%%%%%%%%%%%%%%%%%%%%%%%%%%%%%%%%%%%%%%%%%%%%
\phantomsection
\section*{References}
%\addcontentsline{toc}{section}{References} %Uncomment to add References to content
%%%%%%%%%%%%%%%%%%%%%%%%%%%%%%%%%%%%%%%%%%%%%%%%%%%%%%%%%%%%%%%%%%%%%%%%%%%%%
\begin{description}
%%%%%%%%%%%%%%%%%%%%%%%%%%%%%%%%%%%%%%%%%%%%%%%%%%%%%%%%%%%%%%%%%%%%%%%%%%%%%
\item{} Antia,~H.\,M., Basu,~S., \& Chitre,~S.\,M.
    1998, \mnras\ {\bf 298}, 543
\item{} Barekat,~A., Schou,~J., \& Gizon,~L.
    2014, \aa\ {\bf 570}, L12
\item{} Beck,~P.\,G., Montalban,~J., Kallinger,~T., De\,Ridder,~J., \\ Aerts,~C., Garc\'{i}a,~R.\,A., Hekker,~S., Dupret,~M.-A., et al.
    2012, \nat\ {\bf 481}, 55
\item{} Brandenburg,~A.
    2005, \apj\ {\bf 625}, 539
\item{} Charbonneau,~P., Christensen-Dalsgaard,~J., Henning,~R.,  Larsen,~R.\,M., Shou,~J., Thompson,~M.\,J., \& Tomczyk,~S.
    1999, \apj\ {\bf 527}, 445
\item{} Deheuvels,~S., Garc\'{i}a,~R.\,A., Chaplin,~W.\,J., Basu,~S.,
    Antia,~H.\,M., Appo\-ur\-ch\-aux,~T., Benomar,~O., Davies,~G.\,R., et al.
    2012, \apj\ {\bf 756}, 19
\item{} K\"apyl\"a,~P.\,J., \& Brandenburg,~A.
    2008, \aa\ {\bf 488}, 9
\item{} K\"apyl\"a,~P.\,J., Mantere,~M.\,J., \& Brandenburg,~A.
    2011, \an\ {\bf 332}, 883
\item{} Kitchatinov,~L.\,L.
    2005, Physics Uspekhi {\bf 48}, 449
\item{} Kitchatinov,~L.\,L.
    2013, in: Solar and Astrophysical Dynamos and Magnetic Activity, IAU Symp. 294 (Eds. A.\,G.~Kosovichev, E.~de Gouveia Dal Pino, Y.~Yan, Cambridge Univ. Press), p.399
\item{} Kitchatinov,~L.\,L., R\"udiger,~G., \& K\"uker,~M.
    1994a, \aa\ {\bf 292}, 125
\item{} Kitchatinov,~L.\,L., Pipin,~V.\,V., \& R\"udiger,~G.
    1994b, \an\ {\bf 315}, 157
\item{} Krause,~F., \& R\"adler,~K.-H.
    1980, Mean-Field Magnetohydrodynamics and Dynamo Theory (Oxford: Pergamon Press)
\item{} Lebedinskii,~A.\,I.
    1941, Astron. Zh. {\bf 18}, 10
\item{} Moffatt,~H.\,K.
    1978, Magnetic Field Generation in Electrically Conducting Fluids (Cambridge Univ. Press)
\item{} Obridko,~V.\,N., Sokoloff,~D.\,D., Kuzanyan,~K.\,M.,
    Shel\-ti\-ng,~B.\,D., \& Zakharov,~V.\,G.
    2006, \mnras\ {\bf 365}, 827
\item{} Pipin,~V.\,V., \& Kosovichev,~A.\,G.
     2011, \apj\ {\bf 727}, L45
\item{} R\"udiger,~G.
    1989, Differential rotation and stellar convection: Sun and solar-type stars (New York: Gordon \& Breach)
\item{} Stenflo,~J.\,O.
    1988, Astrophys. Space Sci. {\bf 144}, 321
\item{} Stix,~M.
    1989, The Sun: An introduction (Berlin: Springer--Verlag)
\item{} Thompson,~M.\,J., Toomre,~J., Anderson,~E.\,R., Antia,~H.\,M.,
    Berthomieu,~G., Burtonclay,~D., Chitre,~S.\,M., Chri\-s\-ten\-sen-Dalsgaard,~J., et al.
    1996, Science {\bf 272}, 1300
\end{description}
\end{document}